\documentstyle[prl,aps,epsfig,floats,twocolumn]{revtex} 
\begin{document}
\draft
\twocolumn[\hsize\textwidth\columnwidth\hsize\csname @twocolumnfalse\endcsname
\title{Mean field and corrections for the Euclidean \\
Minimum Matching problem}

\author{Jacques H.~Boutet de Monvel $^*$
and Olivier C.~Martin $^\dag$ }
\address{Division de Physique Th\'eorique,
Institut de Physique Nucl\'eaire, Universit\'e\ Paris-Sud, F--91406
Orsay Cedex, France}

\date{Submitted to Physical Review Letters }
\maketitle
\begin{abstract}
Consider the length $L_{MM}^E$ of the
minimum matching of $N$ points in $d$-dimensional Euclidean space.
Using numerical simulations and the finite size scaling law
$\langle L_{MM}^E \rangle = \beta_{MM}^E(d) N^{1-1/d}(1+A/N+\cdots )$,
we obtain precise estimates of
$\beta_{MM}^E(d)$ for $2 \le d \le 10$.
We then consider the approximation where
distance correlations
are neglected. This model is solvable \cite{MezardParisi_86b}
and gives at $d \ge 2$
an excellent ``random link'' approximation to $\beta_{MM}^E(d)$.
Incorporation of three-link correlations
further improves the accuracy, leading to
a relative error of
$0.4\%$ at $d=2$ and $3$.
Finally, the large $d$ behavior of this expansion in link correlations
is discussed.
\end{abstract}

\pacs{PACS numbers: 75.50.Lk, 64.60.Cn}
]
\narrowtext


There has been a tremendous amount of work on
mean field calculations for disordered systems
in the past 20 years, in part driven
by the exact solution provided by Parisi's replica
symmetry breaking Ansatz. Although his solution was developed
in the context of spin glasses, the formalism has been
extremely useful for understanding other disordered systems.
Generally, one expects mean field
to give exact results as
the dimensionality goes to infinity. One can then
ask whether mean field leads to ``acceptable''
errors for systems of interest, e.g., in three
dimensions,
and whether it is possible to
compute Euclidean corrections to the mean field formulae. Such
computations might correspond to a
$1/d$ expansion for the thermodynamic functions of interest.
In frustrated disordered systems, however, this has turned out to
be intractable. To date,
such Euclidean corrections have been pushed furthest \cite{MezardParisi_88}
for the Minimum Matching Problem (MMP) \cite{PapadimitriouSteiglitz}.
In what follows, we determine the accuracy of the mean field
approximation and the effectiveness of the corrections thereto
in the MMP by comparing with the actual
properties of the $d$-dimensional Euclidean model.
First, we find that the relative error introduced by the
mean field approximation for the zero temperature energy density is less
than $4 \%$ at $d=2$ and $3 \%$ at $d=3$. Second, the inclusion of the
``leading'' Euclidean corrections to the mean field approximation reduces
the error by a factor of about $10$ at $d=2$ and $d=3$
Third, we argue that the large $d$
behavior of systems such as the MMP depends on arbitrarily high order
correlations and is thus beyond all orders of the expansion
proposed by M\'ezard and Parisi \cite{MezardParisi_88}.

Consider $N$ points ($N$ even) and a specified set of
link lengths $l_{ij}=l_{ji}$ separating the
points, for $1 \le i,j \le N$. One defines
a matching (a dimerization) of these points
by combining them pairwise so that each
point belongs to one and only one
pair. Define also the energy or length
of a matching as the sum of the lengths of the links
associated with each matched pair.
The Minimum Matching Problem is the problem of finding
the matching of minimum energy.
One can also consider the thermodynamics of
this system, as proposed by Orland \cite{Orland} and
M\'ezard and Parisi \cite{MezardParisi_85}, by
taking {\it all} matchings
but weighting them with the Boltzmann factor associated with
their energy. Here we
concentrate on the $T=0$ properties because
there is no effective numerical method for extracting
thermodynamical functions in this system, but exact
energy minima can be obtained quite easily for any given
instance. Indeed, the MMP belongs to the algorithmic
class P of polynomial problems, and
there are standard algorithms which solve any instance of
size $N$ using on the order of $N^3$ steps \cite{BallDerigsMBM_Net}.

Physically, one is not interested in the properties of any particular
instance of the MMP; more relevant are typical and ensemble properties
such as the average energy when the
lengths $l_{ij}$ are random variables
with a given distribution. One then speaks of the
stochastic MMP.
There are two frequently used ensembles for the $l_{ij}$,
corresponding to the {\it Euclidean} MMP and the {\it random link} MMP.
In the first, the $N$ random points lie in a $d$-dimensional
Euclidean volume and the $l_{ij}$ are the
usual Euclidean distances between pairs of points.
The points are independent and identically
distributed, so one speaks of a random point problem.
In the second ensemble,
it is the link lengths $l_{ij}$ which are {\it independent}
and identically distributed random variables.
A connection between these two systems was first given by
M\'ezard and Parisi \cite{MezardParisi_86b}: they pointed out that the
one and two-link distributions
could be made identical in the two problems. A consequence
is that the ``Cayley tree''
approximation for the random point and random
link problems are the same.
M\'ezard and Parisi were able to solve the random link MMP using
an approach
based on replicas \cite{MezardParisi_85,MezardParisi_86b}.
One may then consider the random link MMP to be a ``mean field
model'' for the Euclidean MMP. The mean field {\it approximation}
consists of using the thermodynamic functions
of the random link model as estimators for those of the
Euclidean model. This approximation is applicable to all
link based combinatorial optimization problems, such as
the assignment problem and the traveling salesman problem;
hereafter we shall refer
to it as the {\it random link approximation} \cite{CBBMP}.
Finally, M\'ezard and Parisi have shown how to
derive corrections systematically to the random link approximation
using a connected-correlation link expansion. In \cite{MezardParisi_88},
they have computed the leading corrections, these being associated
with the triangle inequality (3-link correlations) in
the Euclidean model.  How accurate are these approximations?
To answer this, we first
give our results for the Euclidean problem, and then
compare with the predictions
of the random link approximation and of the link expansion method.

In the Euclidean MMP, let $L_{MM}^E$ be the energy or length
of the minimum matching. Taking the points to be independent
and uniformly distributed in a unit volume,
Steele \cite{Steele_AP} has shown that
as $N \to \infty$, $L_{MM}^E / N^{1-1/d}$ converges with
probability one to a non-random, $N$-independent constant
$\beta_{MM}^E(d)$. In physics language, this result shows that
$L_{MM}^E$ is self-averaging and that the
zero-temperature energy density has an infinite volume limit when
the density of points is kept fixed.
To date, little has been done to compute $\beta_{MM}^E(d)$.
The best estimates seem to be due to Smith \cite{Smith_Thesis}:
$\beta_{MM}^E(2) \approx 0.312$ and $\beta_{MM}^E(3) \approx 0.318$.
Let us use a systematic procedure \cite{CBBMP}
to obtain $\beta_{MM}^E(d)$ with quantifiable errors.
First, in order to have a well defined dependence on $N$, we have used
the ensemble average, $\langle L_{MM}^E \rangle / N^{1-1/d}$.
Second, in order to reduce corrections to scaling
in the extrapolation to the large $N$ limit, we have placed the points
randomly in the $d$-dimensional unit hypercube with periodic
boundary conditions. This removes surface effects
and empirically leads
to the finite size scaling law
\begin{equation} \label{eq_fss}
 { \langle L_{MM}^E \rangle  \over N^{1-1/d} } =
  \beta_{MM}^E(d) ( 1 + {A(d) \over N} + { B(d) \over N^2} + \cdots).
\end{equation}
Finally, in order to reduce statistical fluctuations,
we have used a variance reduction trick \cite{CBBMP}.
The improved estimator has the effect of reducing the
variance of our estimates by more than a factor $4$,
and thus saves us a considerable amount of computer time.
Using these methods, we have extracted from our numerical data
$\beta_{MM}^E(d)$ and its associated statistical error.
The fits to Eq.(\ref{eq_fss}) are good, with $\chi^2$ values
confirming the form of the finite size scaling law.
The error bars on the extrapolated value $\beta_{MM}^E(d)$
are obtained in the standard way by requiring
that $\chi^2$ increase by one from its minimum.
We find in particular $\beta_{MM}^E(2) = 0.3104 \pm 0.0002$,
and $\beta_{MM}^E(3) = 0.3172 \pm 0.00015$;
values at higher dimensions are given in Table \ref{tab_betas}.
We have checked that these results are
not significantly modified when using another random number
generator to produce the instances, and that the fits are stable
to truncation of the data.

Now we discuss how to use the random link model
to approximate $\beta_{MM}^E(d)$.
For any two points $(i,j)$ placed
at random in the unit $d$-dimensional hypercube, the density
distribution of $l_{ij}$ is given at short distances by
$P_d(l_{ij}=r) = d B_d r^{d-1}$, where
$B_d = \pi^{d/2}/(d/2)!$ is the volume of the $d$-dimensional
ball with unit radius.  If we take the random link model where
link lengths are independent
and have the individual distribution $P_d(l)$, then the Euclidean
and random link MMP have the same one and two-link
distributions \cite{MezardParisi_86b} because two Euclidean distances are
independent. If correlations among three or more link lengths
are weak, then the properties of the two systems
should be quantitatively close.  Thus an
analytic approximation to $\beta_{MM}^E(d)$ is obtained by computing
its analogue $\beta_{MM}^{RL}(d)$ in the random link MMP.
In references \cite{MezardParisi_85,MezardParisi_86b}, M\'ezard and Parisi
solved these random link models under the replica symmetry hypothesis.
They showed further \cite{MezardParisi_87b} that the replica
symmetric solution is stable (at least for $d=1$),
and thus is most likely exact unless a first order phase transition
occurs in this system. Their solution
gives $\beta_{MM}^{RL}(d)$ in terms of a function $G_d$ related
to the probablility distribution of link lengths for matched pairs.
In our Euclidean units their result can be written
\begin{equation} \label{eq_mf_beta}
\beta_{MM}^{RL}(d) = {D_1(d)\over 2} {d\over (1/d)!}
                    \int_{-\infty}^{+\infty} G_d(x) e^{-G_d(x)} dx
\end{equation}
where $G_d$ satisfies the integral equation
\begin{equation} \label{eq_mf_G}
G_d(x) = d \int_{-x}^{+\infty} (x+y)^{d-1} e^{-G_d(y)} dy
\end{equation}
and where \begin{equation}
D_1(d) = \lim_{N\to \infty} \langle L_1 \rangle /N^{1-1/d} = (1/d)!
B_d^{-1/d}
\end{equation}
is the average (rescaled) link length of the nearest neighbor
graph in the limit $N\to \infty$.

\begin{figure}
\begin{center}
\begin{picture}(300,300)
\epsfig{file=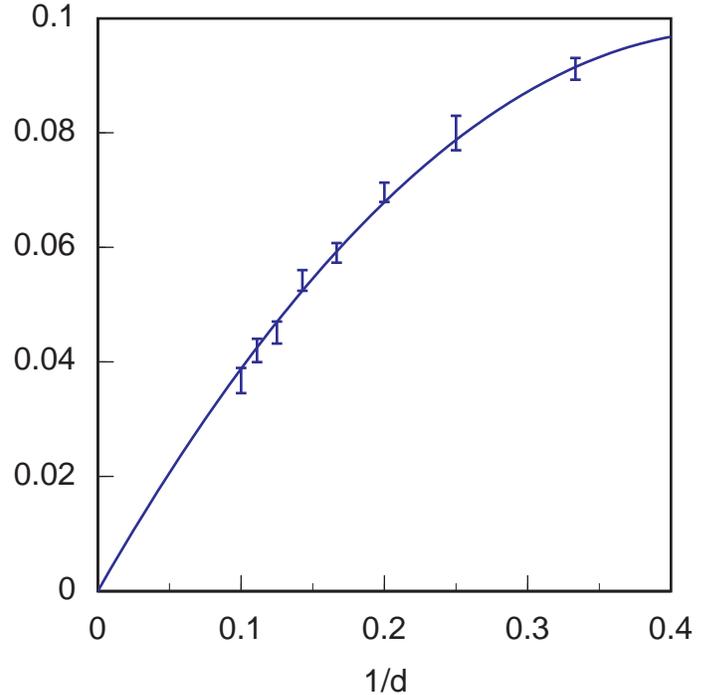}
\end{picture}
\caption{Linear scaling with $1/d$ of the quantity
$d(\beta^{RL}-\beta^E)/\beta^{E}$.}
\label{fig_rl_error}
\end{center}
\end{figure}

Brunetti {\it et al.} \cite{BKMP} have used
direct numerical simulations of these random link models
to confirm the predictions to the level of $0.2 \%$
at $d=1$ and $2$, and we
have done the same to the level of $0.1 \%$ at $1 \le d \le 10$,
giving further evidence that the replica symmetric solution is exact.
{}From the analytical side, solving the integral
equation for $G_d$
leads to $\beta_{MM}^{RL}(1)=\pi^2/24 = 0.4112335$\ldots,
$\beta_{MM}^{RL}(2) = 0.322580$\ldots,
and $\beta_{MM}^{RL}(3)=0.326839$\ldots;
values at higher dimensions are given in Table \ref{tab_betas}.
If we consider $\beta_{MM}^{RL}(d)$ as a
mean field prediction for $\beta_{MM}^E(d)$, the accuracy
is surprisingly good.
Including the trivial value $\beta_{MM}^E(1) = 0.5$,
we see that the random link approximation leads to a relative
error of $17.8 \%$ at $d=1$, of $3.9 \%$ at $d=2$, and of $3.0 \%$ at $d=3$.
Also, the error decreases with increasing dimension.
It can be argued, for the MMP as well as for other link-based
combinatorial optimization problems \cite{CBBMP}, that the
random link approximation
not only has a relative error tending towards $0$ as $d \to \infty$, but that
in fact this error is at most of order $1/d^2$. Given our high
quality estimates, we are able to confirm this
property numerically. In Figure \ref{fig_rl_error} we plot
the quantity $d (\beta_{MM}^{RL}-\beta_{MM}^E)/ \beta_{MM}^{E}$
along with a quadratic fit given to guide the eye.
As expected, the data scales as $1/d$.
Thus the random link approximation gives both the leading and
$1/d$ subleading dependence of $\beta_{MM}^E(d)$. In order
to obtain analytic expressions for these coefficients,
we have derived the $1/d$ expansion for $\beta_{MM}^{RL}$ from
Eqs.(\ref{eq_mf_beta},\ref{eq_mf_G}). We used two methods to do this.
The first, straightforward but computationally lengthy,
consists of setting ${\tilde G_d}(x) = G_d(\tilde x = x/d + 1/2)$
and then writing ${\tilde G_d}(x)$ as a power series in $1/d$.
{}From this we find
\begin{equation} \label{eq_betaRL}
\beta_{MM}^{RL}(d) = {D_1(d)\over 2} (1 + {1-\gamma \over d}
                       + O(1/d^2) )
\end{equation}
where $\gamma = 0.577\ldots$ is Euler's constant.
If, as claimed, the random link approximation gives an
error of order $1/d^2$, Eq.(\ref{eq_betaRL}) gives
an analytic expression for the leading and first subleading
terms in the $1/d$ expansion of $\beta_{MM}^E(d)$.
This claim is strongly supported by the numerical results:
performing a fit of our $\beta_{MM}^E(d)$
values to a truncated $1/d$ series leads to
$0.424 \pm 0.008$ for the coefficient of
the $1/d$ term; this is to be compared to the theoretical
prediction of $1-\gamma = 0.422784\ldots$

\par
We have been able to obtain the next coefficient of the series in
$1/d$ for $\beta_{MM}^{RL}$
by using a second method. We introduce a modified random link model
where the links are shifted and rescaled in such a way that the leading term
of the $1/d$ expansion for this new model is exactly the $1/d$ coefficient for
the initial one \cite{Boutet_Thesis}. In fact it is possible to introduce a
sequence of such ``rescaled'' models, where the $k^{th}$ model is designed to
produce the $1/d^k$ term of the expansion. We have computed the leading terms
predicted by a replica symmetric analysis of these
models for $k=1$ and $2$, from
which we find that the order $1/d^2$ coefficient in Eq.(\ref{eq_betaRL}) is
$\pi^2/12 + \gamma^2/2 - \gamma$.

We now come to the final point of the paper: how well
can one predict $\beta_{MM}^E(d)$ by incorporating
Euclidean corrections to the random link approximation?
It is necessary here to review the work of M\'ezard
and Parisi; for greater detail, we refer the reader to
their article \cite{MezardParisi_88}. They begin with the
partition function $Z$ for an arbitrary stochastic MMP
and write the quenched average for $n$ replicas.
In the Euclidean model, the $l_{ij}$ have three and
higher-link correlations. M\'ezard and Parisi keep
the three-link correlations (arising only when the
three links make a triangle) and neglect higher
connected correlations.
Note that it is not
clear {\it a priori} whether these ``higher order'' terms
are negligeable compared to the three-link term.
The resulting expression for the quenched average becomes
\begin{eqnarray}  \label{eq_Z_EC}
{\overline {Z^n}} = \prod_{j=1}^N \prod_{\alpha=1}^n \big(
\int_0^{2 \pi} {d \lambda_j^{\alpha} \over 2 \pi} ~ e^{i \lambda_j^{\alpha} }
\big)
\nonumber \\
\times ~~ e^{ \sum_{(ij)} {\overline {u_{ij}} } +
\sum_{(ij)(kl)(mn)}^{\prime} {\overline {u_{ij} u_{kl} u_{mn} } }^{~C}  }
\end{eqnarray}
where $u_{ij}$ is a complicated nonlinear function of
the link length $l_{ij}$.
They then compute the limit $N \to \infty$, $n \to 0$
using the saddle point method while assuming that
replica symmetry is not broken.
In the zero temperature limit, just as in the standard random
link model, the saddle point equations can be expressed in
terms of $G_d$, but now $G_d$ satisfies a
more complicated integral equation (Eq.(34) in their paper).
{}From this, one can calculate new estimates for
$\beta_{MM}^E(d)$, which we shall denote $\beta_{MM}^{EC}$,
where $EC$ stands for Euclidean corrections.

\par
We have solved the equations numerically for this modified
$G_d$, and have computed $\beta_{MM}^{EC}(d)$ for $2 \le d \le 10$.
We find $\beta_{MM}^{EC}(2)=0.30915$ and
$\beta_{MM}^{EC}(3)=0.31826$.
The results for $d \ge 4$ are given in Table \ref{tab_betas}.
Comparing with $\beta_{MM}^E(d)$ and
$\beta_{MM}^{RL}(d)$, we see that the new estimates are considerably
more accurate. At $d=2$, the random link approximation leads to
an error of $3.9 \%$; this error is decreased by nearly
a factor $10$
by incorporating these leading Euclidean corrections.
Similarly at $d=3$, the error is reduced from $3.0 \%$ to
less than $0.4\%$. At larger $d$, the error
continues to decrease, but the effect is
less dramatic.

\par
To interpret this last result, consider how the
difference $\beta_{MM}^{EC} - \beta_{MM}^{RL}$
scales with $d$. Using Eq.(\ref{eq_Z_EC}), we see that it
is sufficient to estimate the
size of the $3$-link correction term.
Its dimensional dependence follows that of the
probability of finding nearly equilateral triangles as $d \to \infty$.
Since this probability goes to zero exponentially with $d$,
the $3$-link correlations give tiny corrections
at large $d$ (as confirmed by the numreics), and also the power
series expansion in $1/d$ of $\beta_{MM}^{EC}$ is {\it identical}
to that of $\beta_{MM}^{RL}$. This property continues to hold
if one includes 4, 5, or any {\it finite} number of multi-link correlations
in Eq.(\ref{eq_Z_EC}).
This is due to the fact that the Euclidean and random link graphs
have {\it local} properties that are identical
up to exponentially small terms in $d$. In particular, the statistics
of fixed sized ($N$-independent) loops connecting near neighbors are nearly
identical.

\par
Although this reasoning was given for the MMP,
it applies equally well to
other link-based problems. In such
statistical mechanics systems, if the thermodynamic functions
depend only on the local properties of the (short) link graph,
then the random link approximation applied to the Euclidean system
will have an error which is
exponentially small in $d$.
However, for combinatorial optimization problems such as the MMP,
the assignment problem, and
the traveling salesman problem,
the $N \to \infty$ limit and the $k$-link expansion do not
commute: $k$-link correlations
with $k$ growing with $N$ remain important as $N \to \infty$.
In particular, arbitrarily large loops matter and contribute to the
thermodynamics at order $1/d^2$. In a polymer picture,
we can say that the
random link approximation is exponentially good in the dilute phase, while
it leads to $1/d^2$ errors in the collapsed phase. The power corrections
in this phase are beyond all orders in a $k$-link correlation expansion such
as Eq.(\ref{eq_Z_EC}).

\par
In summary, we have estimated by numerical simulation
$\beta_{MM}^E(d)$, the zero energy density in the
Euclidean Minimum Matching problem at dimensions $2 \le d \le 10$.
We have then computed two analytical estimates for these energy
densities, namely $\beta_{MM}^{RL}(d)$ and $\beta_{MM}^{EC}(d)$.
The first method uses the random link approximation where all
link correlations are neglected.
Using the ``exact'' mean field solution of M\'ezard
and Parisi, we find that even
at low dimensions, the error introduced by this approximation
is small: $3.9 \%$ at $d=2$, $3.0 \%$ at $d=3$,
and $2.0 \%$ at $d=4$. In the second
method, the connected three-link correlations are taken into
account while higher ones are neglected.
Using M\'ezard and Parisi's expressions, we find
that this modification
to the random link model gives excellent predictions
at $d=2$ and $3$, with the error there being
divided by almost $10$ compared to the random link approximation.
This provides a stringent quantitative test of
a systematic expansion which goes beyond uncorrelated disorder variables,
and suggests that even the leading such correction
is enough to get predictions for thermodynamic functions
precise to better than one percent.
Finally, at high dimensions, we have seen that the $k$-link
correlation expansion leads to corrections which vanish exponentially
with $d$; this expansion thus misses important $1/d$ power law corrections
for problems such as the MMP.
This leaves open the determination of the $1/d^2$ term in the expansion of
the constants $\beta_{MM}^E(d)$. We have performed a fit on our data,
imposing the leading and the $1/d$ term to be those of the random link model.
We find the $1/d^2$ coefficient to be very small (smaller than $0.01$ in
absolute value). Clearly, it would be of major interest to obtain an
analytical value for this term.

\par
We are grateful to C. De Dominicis, J. Houdayer,
W. Krauth, M. M\'ezard, and H. Orland
for their interest and suggestions.
JBdM acknowledges a fellowship from the MENESR, and OCM
acknowledges support from the Institut Universitaire de France.
The Division de Physique Th\'eorique is an Unit\'e de Recherche
des Universit\'es Paris XI et Paris VI associ\'ee au CNRS.

\par
{*} Electronic address: boutet@ipno.in2p3.fr
\par
{\dag} Electronic address: martino@ipno.in2p3.fr
\par

\begin{thebibliography}{10}

\bibitem{BallDerigsMBM_Net}
M.~O. Ball and U.~Derigs.
\newblock An analysis of alternate strategies for implementing matching
  algorithms.
\newblock {\em Networks}, 13:517--549, 1983.

\bibitem{Boutet_Thesis}
J.~{Boutet de Monvel}.
\newblock {\em Physique statistique et mod\`eles \`a liens al\'eatoires}.
\newblock PhD thesis, Institut de Physique Nucl\'eaire, Universit\'e Paris-Sud,
  Orsay, 1996.

\bibitem{BKMP}
R.~Brunetti, W.~Krauth, M.~M\'ezard, and G.~Parisi.
\newblock Extensive numerical solutions of weighted matchings: Total length and
  distribution of links in the optimal solution.
\newblock {\em "Europhys. Lett."}, 14(4):295--301, 1991.

\bibitem{CBBMP}
N.~J. Cerf, J.~{Boutet de Monvel}, O.~Bohigas, O.~C. Martin, and A.~G. Percus.
\newblock The random link approximation for the {E}uclidean traveling salesman
  problem.
\newblock {\em J.~Phys. I France}, 7(1):117--136, 1997.

\bibitem{MezardParisi_85}
M.~M\'ezard and G.~Parisi.
\newblock Replicas and optimization.
\newblock {\em "J.~Phys. Lett. France"}, 46:L771--L778, 1985.

\bibitem{MezardParisi_86b}
M.~M\'ezard and G.~Parisi.
\newblock Mean-field equations for the matching and the travelling salesman
  problems.
\newblock {\em "Europhys. Lett."}, 2:913--918, 1986.

\bibitem{MezardParisi_87b}
M.~M\'ezard and G.~Parisi.
\newblock On the solution of the random link matching problems.
\newblock {\em "J.~Phys. France"}, 48:1451--1459, 1987.

\bibitem{MezardParisi_88}
M.~M\'ezard and G.~Parisi.
\newblock The {E}uclidean matching problem.
\newblock {\em "J.~Phys. France"}, 49:2019--2025, 1988.

\bibitem{Orland}
H.~Orland.
\newblock Mean-field theory for optimization problems.
\newblock {\em "J.~Phys. Lett. France"}, 46:L763--L770, 1985.

\bibitem{PapadimitriouSteiglitz}
C.~H. Papadimitriou and K.~Steiglitz.
\newblock {\em Combinatorial Optimization: Algorithms and Complexity}.
\newblock Prentice Hall, Englewood Cliffs, NJ, 1982.

\bibitem{Smith_Thesis}
W.~D. Smith.
\newblock {\em Studies in Computational Geometry Motivated by Mesh Generation}.
\newblock PhD thesis, Princeton University, Princeton, NJ, 1989.

\bibitem{Steele_AP}
M.~Steele.
\newblock Subadditive {E}uclidean functionals and nonlinear growth in geometric
  probability.
\newblock {\em The Annals of Probability}, 9(3):365--376, 1981.

\end{thebibliography}

\begin{table}
\caption{Comparison of the MMP constants for the three models:
Euclidean, random link, and random link including 3-link Euclidean
corrections ($4\le d\le 10$).}
\label{tab_betas}
\begin{tabular}{cccccc}
$d$ & $\beta^E(d)$ & $\beta^{RL}(d)$ &
$d {\beta^{RL} - \beta^E\over \beta^E}$ &
$\beta^{EC}(d)$ & ${\beta^{EC}-\beta^E\over \beta^E}$\\
\hline
4 & 0.3365$\pm$ 0.0003 & 0.343227 & +0.080 & 0.33756 & +0.30\%\\
5 & 0.3572$\pm$ 0.00015 & 0.362175 & +0.070 & 0.35818 & +0.27\%\\
6 & 0.3777$\pm$ 0.0001 & 0.381417 & +0.059 & 0.37849 & +0.21\%\\
7 & 0.3972$\pm$ 0.0001 & 0.400277 & +0.054 & 0.39807 & +0.22\%\\
8 & 0.4162$\pm$ 0.0001 & 0.418548 & +0.045 & 0.41685 & +0.17\%\\
9 & 0.4341$\pm$ 0.0001 & 0.436185 & +0.042 & 0.43485 & +0.17\%\\
10 & 0.4515$\pm$ 0.0001 & 0.453200 & +0.037 & 0.45214 & +0.14\%\\
\end{tabular}
\end{table}

\bigskip
\end{document}